\documentclass[useAMS,usenatbib,usegraphicx]{mn2e}
\usepackage{color}
\title[Machine-assisted discovery of relationships in astronomy]{Machine-assisted discovery of relationships in astronomy}
\author[M. J. Graham et al.]{Matthew~J.~Graham,$^1$\thanks{E-mail:mjg@caltech.edu}, S.~G.~Djorgovski,$^1$  Ashish~A.~Mahabal,$^1$ 
\newauthor
Ciro~Donalek,$^1$ and Andrew~J.~Drake$^1$\\
$^{1}$California Institute of Technology, 1200 E. California Blvd, Pasadena, CA 91125, USA}
\begin{document}

\date{Accepted . Received ; in original form}

\pagerange{\pageref{firstpage}--\pageref{lastpage}} \pubyear{2011}

\maketitle

\label{firstpage}

\begin{abstract}
High-volume feature-rich data sets are becoming the bread-and-butter of $21^{\mathrm st}$ century astronomy but present significant challenges to scientific discovery. In particular, identifying scientifically significant relationships between sets of parameters is non-trivial. Similar problems in biological and geosciences have led to the development of systems which can explore large parameter spaces and identify potentially interesting sets of associations. In this paper,  we describe the application of automated discovery systems of relationships to astronomical data sets, focussing on an evolutionary programming technique and an information-theory technique. We demonstrate their use with classical astronomical relationships - the Hertzsprung-Russell diagram and the fundamental plane of elliptical galaxies. We also show how they work with the issue of binary classification which is relevant to the next generation of large synoptic sky surveys, such as LSST.  We find that comparable results to more familiar techniques, such as decision trees, are achievable. Finally, we consider the reality of the relationships discovered and how this can be used for feature selection and extraction.
\end{abstract}


\section{Introduction}

The rate of scientific discovery in astronomy has traditionally been tied to the amount of data available.  
The advent of digital astronomy with modern detectors and computational resources, e.g., databases,
has changed this. Although more data in the past two decades has allowed us to discover the cosmic web, dark energy, and exoplanets, the vast majority of such low-hanging fruit have now been found. 
The new challenge is growing data complexity. The era of data-intensive astronomy promises a vastly more thorough exploration of parameter space but the discovery of new scientifically significant relationships is equally more complicated and daunting when faced with overwhelming data dimensions and volumes.

Given a highly complex data set, such as SDSS, with hundreds of parameters for each object, and sufficient numbers of objects - hundreds of millions or more - to provide a fair and representative covering of parameter space, one will uncover many significant relationships - linear, nonlinear, functional, structural - between pairs, triplets and groups of parameters. However, only a small fraction of these will be truly causative, the result of some valid underlying astrophysical process or processes, and identifying these is non-trivial. In fact, \cite{cew12} have shown that identifying the underlying dynamical equations from any amount of experimental data, however precise, is a provably computationally hard (NP-hard\footnote{In computational complexity theory, a problem that is solvable in polynomial time by a nondeterministic Turing machine is an NP (nondeterministic polynomial time) problem. NP-hard problems are those which are at least as hard as any NP-problem, e.g., given a set of integers, does any non-empty subset of them add up to zero?}) problem.

The framework of astroinformatics, combining astronomy, applied computer science and information technology, has arisen to contend with this computational intractability. At its core are sophisticated data mining and multivariate statistical techniques which seek to extract and refine information from highly complex data sets (see \cite{bb10} for an overall review of different techniques in astronomy, \cite{br11} for those specific to the time domain and the IVOA KDDIG web pages\footnote{International Virtual Observatory Alliance Knowledge Discovery in Databases Interest Group:  http://www.ivoa.net/cgi-bin/twiki/bin/view/IVOA/IvoaKDD} for general material related to this). This includes identifying unique or unusual classes of objects, estimating correlations, and computing the statistical significance of a fit to a model in the presence of missing or bounded data, i.e., with lower or upper limits, as well as visualizing this information in a useful and meaningful manner. However, the nature of these methodologies is, at best, semi-automated with focused application in particular regions of discovery space rather than allowing an unbounded exploration of what might be there. 
 
In recent years a number of approaches have been presented in the general scientific literature that seek to redress this, e.g., \cite{oliver04}, \cite{eureqa09}, \cite{sparkes10}, \cite{mic11}. Discovery-based science employs cutting edge data mining techniques for automated hypothesis forming and automated theorem proving. Many of these tend to have originated within the context of systems biology (out of association analysis) where researchers are attempting to identify and derive universal relationships in biological systems akin to those which seem to exist in physical ones, although there is prior art in computer science, particularly within the area of genetic programming (\cite{koza92}). 

These methods are also similar in scope to various feature selection and extraction and dimensional reduction techniques, such as principal component analysis (e.g., \cite{pca92}) and self-organizing maps (\cite{som}), which attempt to counter the ``curse of dimensionality'' by reducing high dimensional data to lower more manageable dimensions whilst preserving meaningful structures within them. Nonetheless these so-called automated discovery methods are applicable to any general data set and especially to those with many variables, such as arise in economics, climate science, sensor networks or any field advocating an informatics-based approach. 

In this work, we describe the application of automated discovery systems of relationships to astronomical data. We have focused in particular on two types of approach - those that seek to identify general connections (correlations) between particular parameters in a data set and those that try to formulate a specific functional relationship between parameters. These may be considered representative of the type of mapping of discovery space that has so far been attempted. A common complaint of data mining techniques is that they usually follow a ``black box'' approach - the data goes in and the answer comes out but there is no real understanding of how one led to the other. We hope to show that automated discovery systems are also more translucent if not actually transparent and allow some deconstruction of the methodology to understand what is going on inside. This is particularly important if their discoveries are to be scientifically validated, i.e., a particular relationship is not only statistically significant but also stems from a (new) non-trivial underlying cause.

It should be noted that although these discovery tools are labelled as automated, they are actually employed as part of a collaborative human-machine discovery process. In data-intensive problems, not only is data processing and analysis automated but also necessarily, given the data volumes and dimensions, the first levels of data interpretation. The human expert now validates machine-generated hypotheses rather than attempting to formulate them themselves. We still make discoveries, but as the complexity of data increases, we need machine intelligence to help us guide towards an insight. 

This paper is structured as follows: in section 2, we will describe the two specific techniques we are applying whilst in section 3, we will present a number of different astronomical contexts in which these have been applied - these both attempt to mimic or recreate past discoveries as well as find new ones. 
We will analyze and discuss our results in sections 4 and 5 and present our conclusions in section 6.

\section{Automated discovery systems}

The methods we are applying in this paper will probably be unfamiliar to many astronomers and so, in this section, we will introduce some of the pertinent terminology and formalism related to them. 

\subsection{Maximal information coefficient}

The maximal information coefficient (MIC; \cite{mic11}) aims to be the 21st-century equivalent of the Pearson correlation coefficient (\cite{speed11}) but it goes beyond just expressing linear associations and can quantify general associations between variables. It is based on the mutual information between two random variables, $A$ and $B$:

\begin{equation}
\mathrm{MI}(A, B) = \sum_{a \in A} \sum_{b \in B} p(a,b) \log \left( \frac{p(a,b)}{p(a)p(b)} \right)
\end{equation}

\noindent
where $p(a)$ and $p(b)$ are the marginal probability mass functions of $A$ and $B$ and $p(a,b)$ is the joint probability mass function respectively. 

Now consider a partitioning of a data set, $D$, of ordered pairs, $\{(a_i, b_i), i = 1, \ldots, n\}$, into an $x$-by-$y$ grid, $G$, such that there are $x$ bins (of variable size) covering $a$ and $y$ bins (also of variable size) spanning $b$ respectively. The probability mass function of a particular grid cell is clearly proportional to the number of data points falling inside that cell and so, for a given $(x, y)$, there will be a maximal mutual information. We can then construct a characteristic matrix $M(D)$ whose elements:

\[
M(D)_{x,y} =  \frac{\max(\mathrm{MI})}{\log \min \{x,y\}}
\] 

\noindent
are the highest normalized mutual informations achieved by any of the corresponding $x$-by-$y$ grids. The maximal information coefficient is then defined to be the maximum value in $M$, such that $xy <  C$ where $C$ is a function of the sample size and represents the maximal grid size considered. Too high a value for $C$ can lead to non-zero scores even for random data because each data point gets its own cell, while setting it too low means that only simple patterns are considered. \cite{mic11} found empirically that $C = n^{0.6}$ provides a satisfactory limit:

\[
\mathrm{MIC}(D) = \max_{xy < C(n)} \left\{M(D)_{x,y} \right\}
\]

The behaviour of MIC is that it tends to 1 for all never-constant noiseless functional relationships and to 0 for statistically independent variables. Moreover, MIC - $r^2$, where $r$ is the Pearson correlation coefficient, is an indicator of a nonlinear relationship between two variables: as $r$ is a measure of linear dependence, the statistic MIC - $r^2$, is near to 0 for linear relationships and large for nonlinear relationships with high values of MIC. Other measures involving MIC and M (the characteristic matrix) can also indicate deviations from monotonicity, the degree to which the data set appears to be sampled from a continuous function and the complexity of the association, as different relationship types give rise to characteristic matrices with different properties.

The statistical significance of a MIC value can be determined from comparison of a real value against a set of values from $1/\alpha -1$ surrogate data sets where $\alpha$ is the probability of false rejection. Because MIC is a rank-order statistic, the uncorrected p-value (essentially the one-tailed p-value for this statistic; when multiple hypotheses (many parameters) are being tested, a corrected value must be used to mitigate false positives) of a given MIC score under a null hypothesis of statistical independence depends only on the score and on the sample size of the relationship in question and not on the specific relationship being tested. Precomputed uncorrected p-values are available for different sample sizes\footnote{http://www.exploredata.net/Downloads/P-Value-Tables}.

\begin{figure}
\centering
\includegraphics[width=3.5in]{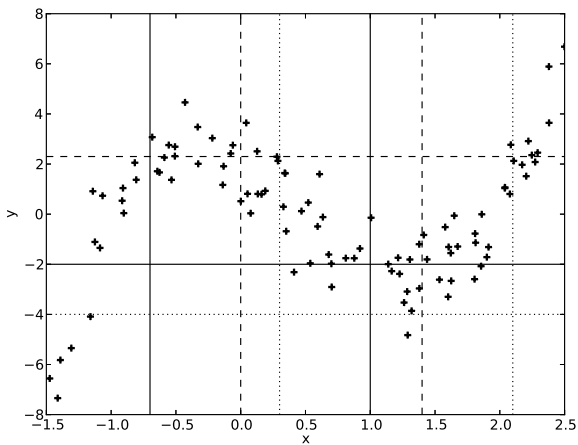} 
\caption{This shows three separate $3 \times 2$ grid partitions of a data set of 100 points randomly selected from a cubic relationship. The mutual information for each grid is: solid line: 0.059, dashed line: 0.044 and dotted line: 0.023 respectively.
}
\label{micexample}
\end{figure}

To illustrate this statistic, consider a data set of 100 points randomly selected from a cubic relationship ($y = 2x^3-3x^2-3x+2$, $x \in [-1.5, 2.5]$) plus a unitary Gaussian noise term, i.e., a Gaussian about the $y$-value with $\sigma = 1$. We can partition this data set into a set of $3 \times 2$ grids (the maximum grid size is set to $xy < 15.8$ for this data) of variable size (see Fig.~\ref{micexample}). Each grid has a mutual information associated with it and for a given partition configuration, e.g., $3 \times 2$, there will be a maximal mutual information. The maximal (normalized) mutual information across all configurations (44 in this case) is the maximal information coefficient. This data set has a statistically significant MIC of 0.836 compared to a linear regression coefficient of just 0.068. It also has high values for the measures indicating nonlinearity (0.831) and  functionality (0.836) and moderate non-monotonicity (0.427).

\subsection{Symbolic regression}

Symbolic regression is the task of finding a function, in symbolic form, that fits a finite sample of data. 
The most efficient approach employs a genetic algorithm-based search (\cite{koza92}) of the space of mathematical expressions to determine the best-fitting functional form. Successive generations of formulae are specified in terms of a (user-defined) mathematical alphabet of atomic building blocks, such as algebraic and boolean operators, analytical function types - trigonometric, exponential/logarithmic, power laws, etc., and state variables, which keeps the search tenable. Its advantage over more standard regression methods is that the search process works simultaneously on both the model specification problem (the form of the fitting equation) and the problem of fitting coefficients.

Eureqa\footnote{http://creativemachines.cornell.edu/eureqa} (now also called Formulize) (\cite{eureqa09}) is a software tool which employs symbolic regression to describe a data set by identifying the simplest mathematical formulae which could describe the underlying mechanisms that produced the data. The tool works from the numerical partial derivatives of each pair of variables in the input data set and uses an evolutionary algorithm to explore this partial differential metric space for non-trivial invariant quantities, looking for predicted partial derivatives that best match the numerical ones:

\[
\left.\frac{\Delta y}{\Delta x} \right|_{D_i} \simeq \left.\frac{\delta y}{\delta x} \right|_{f(x_i, y_i)} = \left. \frac{\delta f}{\delta x} \right/ \frac{\delta f}{\delta y}
\]

\noindent
where $f(x_i, y_i)$ is one of the candidate functions. The search continues until some stopping criterion -- time elapsed, goodness-of-fit, confidence of fit (maturity and stability), etc. -- is met. 
The output is then an ordered list of final candidate analytical expressions on the accuracy-parsimony Pareto front, i.e., the tradeoff between the most optimal (best-fitting according to some criteria) and complexity. Each mathematical operation in an expression has a numerical value (cost) associated with it, e.g., addition = 1, exponentiation = 4, and the complexity of a formula is defined here as the sum of these values. A high-order polynomial could therefore be more complex than a straightforward exponential or trigonometric function.

When comparing and optimizing solutions, Eureqa employs a user-defined error metric. A number of different measures are available and the nature of the data can help determine which is the most appropriate, for example, minimizing the mean of the squared residual errors is suitable for normally distributed noise whereas the logarithmic error is better for many outliers. Data can also be weighted according to some prescription, although the importance of particular variables can always be explicitly stressed in the definition of the equation form being searched for. There are, too, various types of data preprocessing operations available, familiar to data mining, such as normalization, outlier rejection and missing value handling.

The results of symbolic regression, i.e., the expressions identified by Eureqa, are the best (non-trivial) mathematical descriptions of the data. Their interpretation and physical validity, however, remain an exercise for the human expert, who may take them at face value or decide to cross check them ("prove them") using other techniques.

\section{Experiments}

In this section, we report on a number of automated discovery experiments we have carried out with different representative astronomical data sets. A number of different options are available, depending exactly on how you want to measure the whole process. It should be noted that in applying our techniques, we are not simply fitting a set of formulae to data but that the respective discovery methods decide which variables to use and in what functional relationship and then find the optimal coefficients and measures of fit. The two methods are also sufficiently different that it is interesting to compare their findings relative to each other.

\subsection{The Hertzsprung-Russell diagram}
The Hertzsprung-Russell (HR) diagram is the quintessential representation of physical relationships associated with different stages of stellar evolution. The original plot of magnitude vs. temperature can be considered as the two-dimensional PDF, $P(M, T_{eff})$; more modern versions also incorporate metallicty and surface gravity giving a four-dimensional PDF, $P(M, T_{eff}, [M/H], \log g)$ - which constrains all of its arguments. The parameterization of these relationships in terms of observable and non-observable stellar quantities expressed as a function of the observable color is an open problem in astronomy, e.g., \cite{wh03}, \cite{zaninetti08}. This is particularly relevant for the next generation of large photometric surveys, e.g., LSST, where spectroscopy of every source is not feasible. Note that \cite{liu12} describe a related problem of inferring the astrophysical parameters of stars from Gaia spectrophotometry.

\begin{figure}
\centering
\includegraphics[width=3.5in]{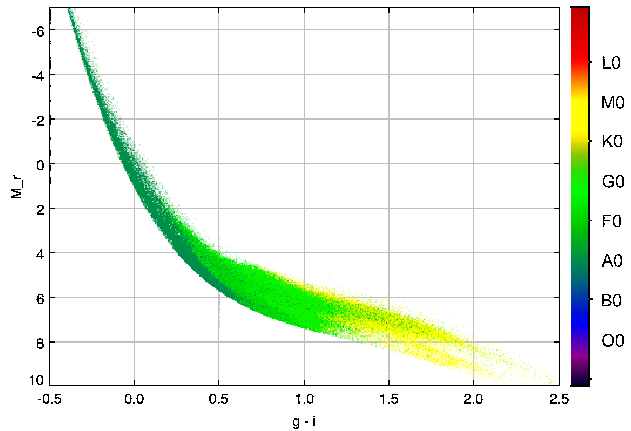} 
\caption{This shows the Hertzsprung-Russell diagram for the SEGUE data using photometric parallax to determine absolute magnitude.}
\label{seguehr}
\end{figure}

\begin{figure}
\centering
\includegraphics[width=3.5in]{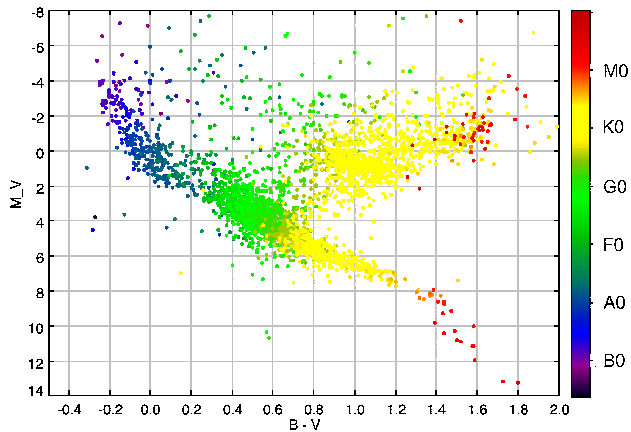} 
\caption{This shows the Hertzsprung-Russell diagram for the 3865 stars in SIMBAD with parallax, effective temperature, surface gravity and metallicity values. The color coding is according to spectral type , broadly defined as: blue - O, B, A; green - F, G; yellow - K; red - M.}
\label{fig1}
\end{figure}

Unfortunately, prior to the availability of the Gaia data, there is no single large stellar data set which offers both accurate distances and physical parameters for a representative sampling of the HR space. Hipparcos has reliable distances but no intrinsic parameters, such as $T_{eff}$ or $[M/H]$.  RAVE DR3 (\cite{rave}) and SEGUE (\cite{segue}) both offer spectroscopically-determined parameters ($T_{eff}, g, [M/H]$) but lack distance information -- RAVE DR3 shares only 685 objects with Hipparcos and SEGUE none. A photometric parallax relationship has been defined for SEGUE based on stellar metallicity and color (\cite{photpar}) but the corresponding HR diagram shows only a main sequence (see Fig.~\ref{seguehr}). For a relatively complete coverage of the parameter space, we have therefore constructed a data set consisting of all stars in SIMBAD with a quoted parallax, effective temperature ($T_{eff}$), surface gravity ($g$), and metallicity $[M/H]$).  Fig.~\ref{fig1} shows the HR diagram for this data set of 3865 stars. 

\begin{figure}
\centering
\includegraphics[width=3.5in]{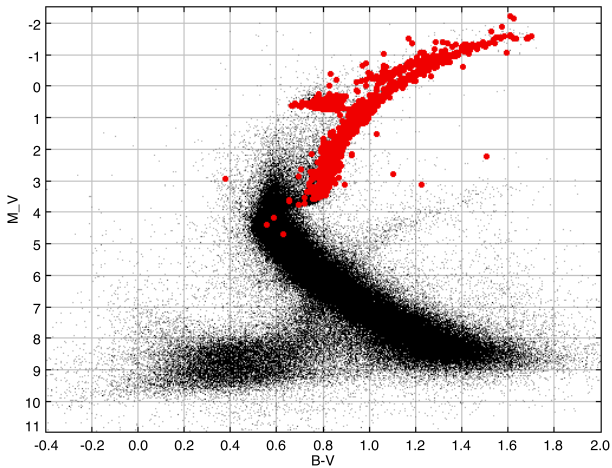}
\caption{This shows the Hertzsprung-Russell diagram for the stars in 47 Tuc taken from Bergbusch \& Stetson (2009). The red points indicate those objects for which spectroscopically-determined parameters are available (Lane el al. (2011)), which are limited to those which have turned off the main sequence.}
\label{47tuchr}
\end{figure}

As a comparison, we have also considered the distribution of parameters for stars from a single globular cluster. 47 Tuc is one of the most studied globular clusters: it is comparatively near, one of the more massive (and hence populous) clusters, and it is relatively metal rich. \cite{lane11} give stellar parameters for 1992 stars in 47 Tuc but only $V$ and $I$ magnitudes. However, \cite{bergbusch09} have measured $B$, $V$, and $I$ for $\sim 200000$ stars in 47 Tuc, giving us a final data set of 1739 stars with stellar parameters: $T_{eff}, g, [m/H]$ (uncalibrated metallicity), $[\alpha/Fe], \xi$ (microturbulence) and $V_{rot}$ (rotational velocity) and $B$ and $V$ magnitudes (see Fig.~\ref{47tuchr} for its HR diagram). 

We ran Eureqa on both data sets to see if we could recover a suitable formulation of the HR relationships, specifically instructing the code to look for formulae of the form:

\[
M_V = f(B-V, g, T_{eff}, [M/H])
\]

\noindent
in the first case (general HR) and:

\[
M_V = f(B-V, g, T_{eff}, [m/H], [\alpha/Fe], \xi, V_{rot})
\]

\noindent
in the second (47 Tuc). Although these formulations are based on prior knowledge of what the dependent variables are and also what data is available, symbolic regression incorporates feature selection and so will only use a subset of the most relevant variables, in this case those which persist in successive generations of calculations in the evolutionary algorithm, rather than all available variables (see section~3.3 for an explicit demonstration of this). We also consider the choice of variables more in section~4.

We use a set of mathematical building blocks restricted to: constants, basic operators (+, -, *, /), exp(), log(), x$^n$. We employed an R$^2$ goodness-of-fit error metric - Eureqa attempts to maximize this quantity in its fits - and selected an 80:20 split of the data in terms of test set and validation set. Data with any missing values were ignored (other options are available) and no weighting was used for any parameter in terms of its error as the heterogeneity of the data means that not every value has an error associated with it.
27 CPU-hour runs (taking 1.5 hours on 18 cores) produced a number of formulae of varying complexity and correlation coefficients of around 0.85 red for both data sets (see Table~\ref{table1}). We also ran it for the SIMBAD data set restricting the formulae to just power law expressions (no exp() or log() operators).

We shall consider the results obtained for the SIMBAD data set first.
To validate the results and test against overfitting, i.e., the formulae are actually describing random errors or noise in the data instead of any underlying relationship, we determined the median absolute error for each formula when applied to the RAVE DR3 and SEGUE data sets mentioned above. Johnson B and V magnitudes were derived from the SEGUE data using the transformation equations of \cite{lupton05} and an absolute V magnitude determined using the inferred parallax relating apparent r magnitude and absolute r magnitude calculated using the photometric parallax method of \cite{photpar}. Any systematic errors that these transformations may introduce can be estimated from a plot of $M_V$ vs. $T_{eff}$ for the SEGUE data compared to the SIMBAD data. Since $T_{eff}$ is calculated spectroscopically, any photometric offset will show in the relative positions of the main sequences of the two data sets. 

\begin{figure*}
\centering
\includegraphics[width=3.4in]{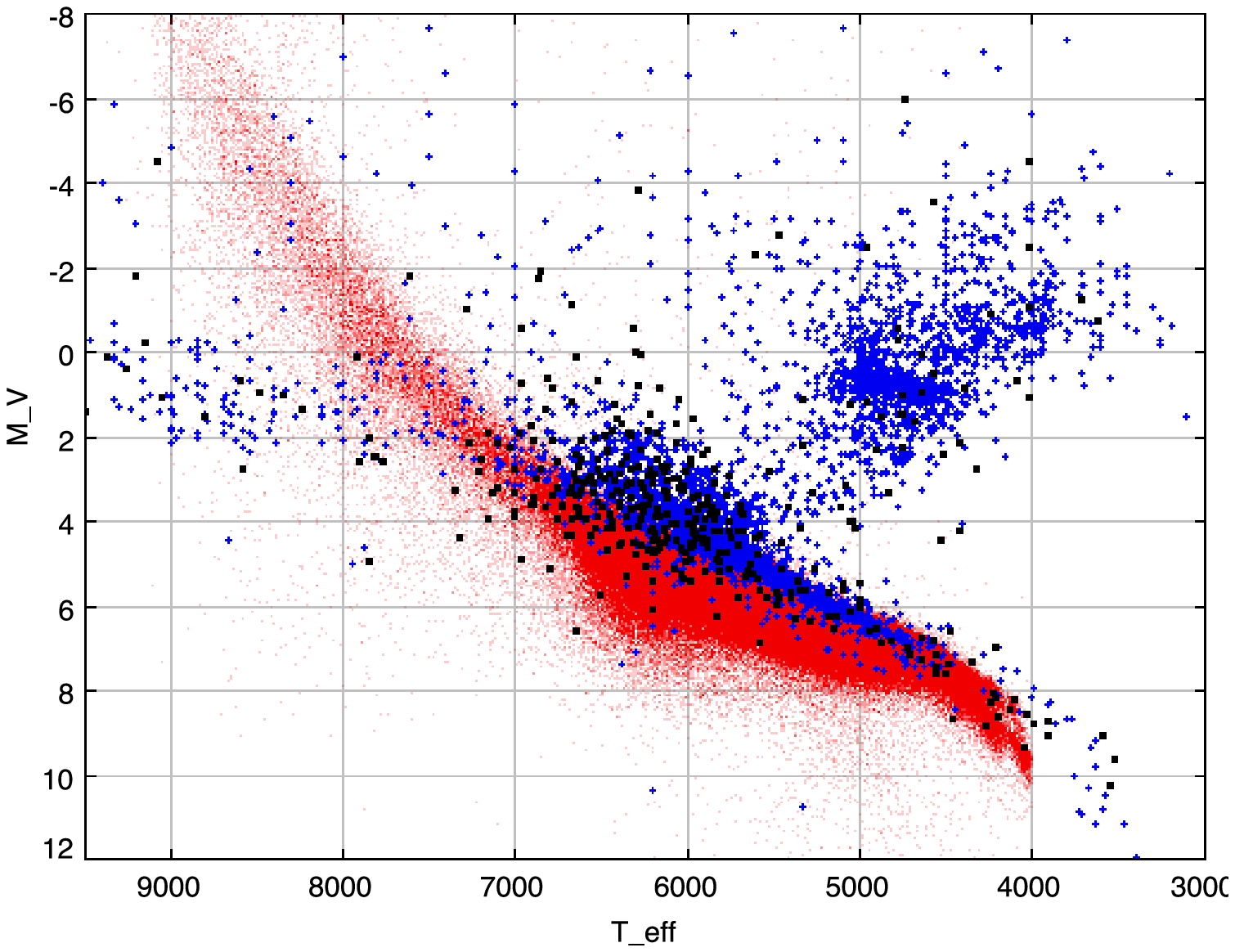} 
\includegraphics[width=3.4in]{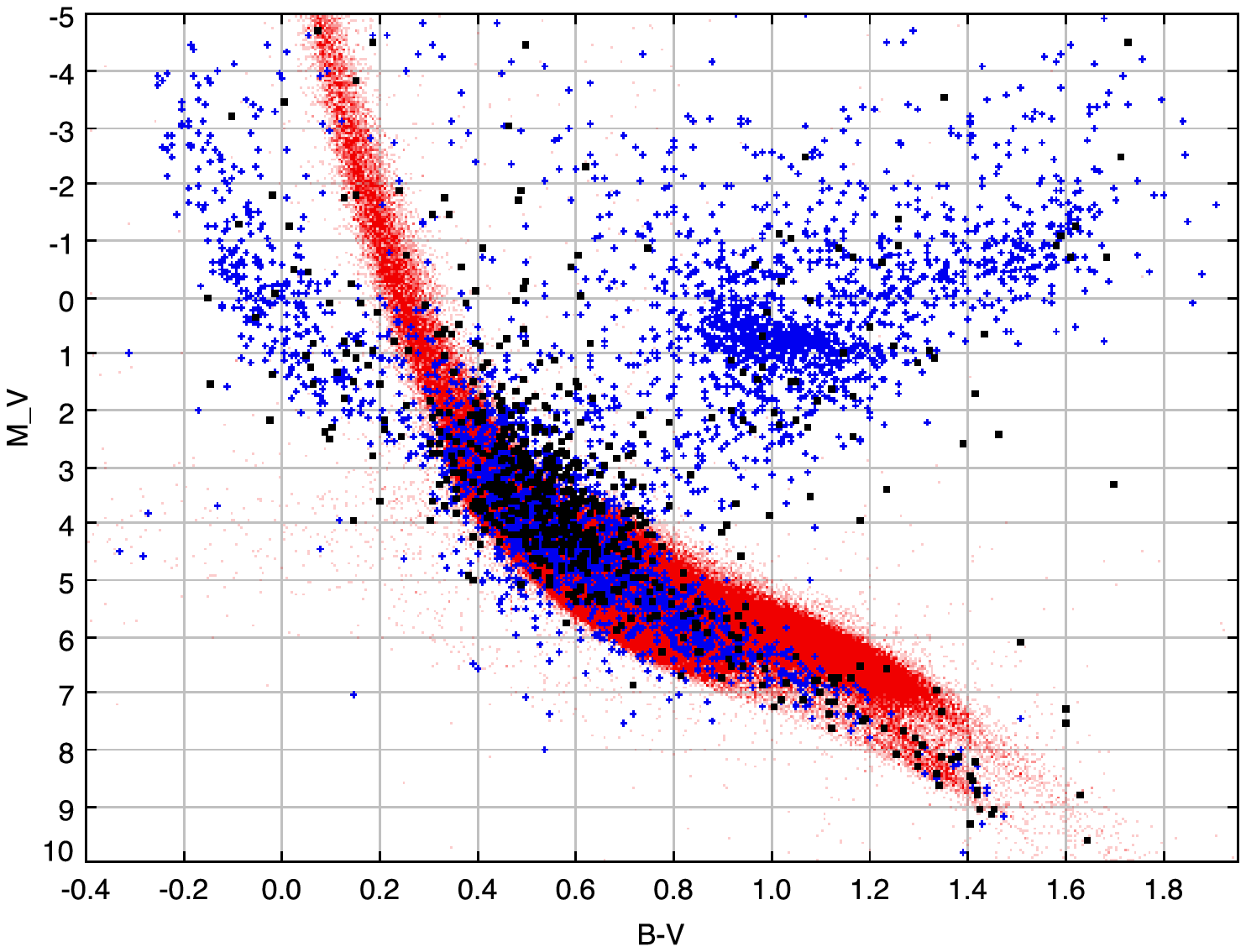}
\caption{This shows the relative positions of the main sequences of SEGUE (red), SIMBAD (blue) and RAVE DR3 (black) data sets. In the left plot, the SIMBAD and RAVE data sets agree well but there is an offset of the SEGUE main sequence relative to the other two introduced by errors in the photometric transformation and parallax methods applied to it. Linear fits to the main sequences estimates the offset as $\Delta M_V = -1.112$. The right plot shows the agreement between the data sets when this offset and a similarly estimated color offset of $\Delta (B-V) = -0.041$ is applied to the SEGUE data. }
\label{mvteff}
\end{figure*}

The left plot in Fig.~\ref{mvteff} shows good agreement between the SIMBAD and RAVE DR3 data sets with an offset of the SEGUE data relative to the other two. This offset can be estimated from the difference between linear fits to the main sequences of both SIMBAD and SEGUE data sets defined between the regions of $T_{eff} = 5000$ and $T_{eff} = 6500$ and we find a value of $\Delta M_V = -1.112$ for SEGUE. A similar procedure can be performed with plots of $(B-V)$ vs. $T_{eff}$ to estimate any systematic errors in the color and we find a value of $\Delta (B-V) = -0.041$ for SEGUE. The right plot in Fig.~\ref{mvteff} shows the agreement between the three data sets when the offsets have been applied. 

Table~\ref{table1} gives the median absolute difference (MAD) between the ``measured'' absolute magnitude and the estimated value for each formula when applied to the SIMBAD, SEGUE and RAVE data sets. For comparison, we also computed the MAD between the observed data and the values derived from the semi-analytical formulae of \cite{zaninetti08} relating $M_V$ and $(B-V)$ for each data set (note that Zaninetti's other formulae relating mass, radius and luminosity to $(B-V)$ all derive from these), although those are only defined over the range $-0.33 < (B - V) < 1.80$ and are stellar luminosity class dependent, with separate relationships for main sequence, giant, supergiant and white dwarf stars. 

It is worth bearing in mind when looking at these results that the various functional relationships that this approach finds are, in some statistical sense, the optimal descriptors of the data - they are phenomenological. Their physical interpretation, however, remains the purview of the human scientist. 
This method aims to identify all potentially interesting, significant relationships without any preconceived bias, e.g., due to some established notion of what should actually be there. 

The results for the SIMBAD and SEGUE data sets are broadly consistent suggesting that the found formulae provide a good description of the variable relationships in the data but do not overfit it. It should be not surprising that the Zaninetti formulae give better results for the SEGUE data set since it essentially just consists of a main sequence and uses that class specific result. The Eureqa results are derived for a range of luminosity types and so give a better broader fit but not necessarily for specific luminosity classes. The poor performance on the RAVE DR3 data set can be largely attributed to the errors on the parallax value (the mean value is 7.63 mas with a mean error of 1.91 mas) and thus the absolute magnitude (54\% of the objects have $\sigma_{M_V} > 1$). If we restrict the analysis to those stars with $\sigma_{M_V} < 1$, we find that the MAD values drop to $\sim 1$ for the Eureqa formulae and $0.6$ for the Zaninetti formula.

\begin{table*}
\caption{The best-fitting formulae found by Eureqa to describe the SIMBAD HR diagram. The goodness-of-fit value ($R^2$) is determined from the 20\% of the data set defined to be the validation set. The error value for each data is the median absolute error when applying the fit. The first three formulae are free to include exponential, logarithmic and power law components  whilst the second are restricted to just power law expressions. }
\label{table1}
\begin{tabular}{lccccc}
\hline
Function ($M_V = f(...)$) & Complexity & $R^2$ & SIMBAD & RAVE DR3 & SEGUE \\
\hline
$0.374 g^2 - 4.34 \times 10^{-4} T_{eff}$ & 9 & 0.823 & 0.569 & 2.104 & 0.932 \\
$(B - V)(1 - [M/H]) + \frac{3.73 \times 10^{4} g}{8396 + T_{eff}} - 7.62 $& 16 & 0.867 & 0.432 & 1.584 & 0.669 \\
$g((B - V) + 2.33) + \frac{6.22 (B-V)}{\exp((B-V)g)} - \log(T_{eff}) - 0.645[M/H] $& 30 & 0.889 & 0.409 & 1.765 &  0.644 \\ 
\hline
$\frac{2133 g^2}{T_{eff}} - 2.48 $& 10 & 0.822 & 0.435 & 1.567 & 0.891 \\
$\frac{2255 g^2}{366 + T_{eff}} - 1.94 - 9.56 \times 10^{-5} T_{eff} - 0.348[M/H] $& 20 & 0.842 & 0.408 & 1.488 &  0.799\\
$\frac{0.641 + 0.791[M/H]}{1.18 - (B - V) - g} + \frac{14000 + 14720g}{6137 + T_{eff} - 917g} - 5.17$ & 29 & 0.868 & 0.401 & 1.528 & 0.776\\ 
\hline
Zaninetti & - & - & 0.597 & 0.822 &  0.504 \\ 
\hline
\end{tabular}
\end{table*}

\begin{table*}
\caption{The best-fitting formulae found by Eureqa to describe the 47 Tuc HR diagram and the median absolute error obtained when applying the fit to the data.}
\label{47tuctab}
\begin{tabular}{lcccc}
\hline
Function ($M_V = f(...)$) & Complexity & $R^2$ & 47 Tuc & SIMBAD PMS\\
\hline
$0.685 + \frac{g}{(B-V)} - \xi$ & 8 & & 0.283 & -\\
$\frac{g - [m/H] - 2.39 [\alpha/Fe]}{B-V} - \xi$ & 12 & & 0.260 & - \\
$13.2 + (0.002 T_{eff} -1.62 g -8.42(B-V)) [m/H] - 12.8 (B-V) $ & 23 &  & 0.239 & 0.990 \\
\hline
Zaninetti & - & - & 0.773 & 0.745 \\ 
\hline
\end{tabular}
\end{table*}

We can also constrain the Eureqa algorithm to use those formulae which contain particular variables or terms: for example, a number of the solutions in both sets of formulae contain a $g^2$ term. A set of formulae derived with these limitations have similar MAD values as for the more generic power law. 

The relationships found for the 47 Tuc data set are more specific since they only cover post-main sequence (PMS) stars. Table~\ref{47tuctab} shows that they fare much better than the Zaninetti formula on this data. We also note that the formula include dependencies on parameters related to convection phenomena in stellar atmospheres as would be expected for PMS stars. The metallicities used in the fitting formulae, $[m/H]$,  are the uncalibrated ones determined by the RAVE pipeline in \cite{lane11} - the uncertainties are 0.1 dex. To compare the fits on PMS stars from the SIMBAD data set, we need to replace $[m/H]$ with an equivalent expression in terms of $[M/H]$. \cite{zwitter08} give a calibration equation for RAVE-derived metallicities: 

\[
[M/H] = 0.938 [m/H] + 0.767[\alpha / Fe] - 0.064 \log g + 0.404
\]

\noindent
but note that $[\alpha/Fe]$ has a typical recovery error of up to 0.15 dex and only spans 0.4 dex in range. Thus, although the SIMBAD data have no measured $[\alpha/Fe]$, we can assume a mean value of 0.2 for use in determining uncalibrated metallicities with reasonable accuracy. We note that the SIMBAD PMS data also shows greater intrinsic scatter than the 47 Tuc data.

\begin{table*}
\caption{The MIC measures between the variables used to define the Hertzspung-Russell diagram for the SIMBAD and 47 Tuc data sets. For these data sets, a value of MIC $> 0.41$ for SIMBAD and MIC $> 0.17$ for 47 Tuc is significant at the $10^{-4}$ level respectively. This also illustrates the $n \times (n-1) / 2$ nature of the output for a data set of $n$ variables.}
\label{ddmic}
\begin{tabular}{llllllll}
\hline
Variable pair & MIC & Nonlinearity & Non-monotonicity & Functionality & Complexity & Linear regression\\
\hline
\multicolumn{7}{l}{\bf{SIMBAD}} \\
$(B-V)$ vs. $T_{eff}$ & 0.82 & 0.40 & 0.02 & 0.82 & 7.11& -0.65 \\
$M_V$ vs. $g$ & 0.63 & 0.04 & 0.05 & 0.63 & 7.11 & 0.77 \\
$M_V$ vs. $T_{eff}$ & 0.54 & 0.48 & 0.08 & 0.54 & 7.11 & -0.24 \\
$g$ vs. $T_{eff}$ & 0.52 & 0.46 & 0.03 & 0.52 & 7.11 & 0.24 \\
$M_V$ vs. $(B-V)$ & 0.49 & 0.47 & 0.09 & 0.48 & 7.11 & -0.12 \\
$(B-V)$ vs. $g$ &0.46 & 0.14 & 0.0 & 0.46 & 7.11 & -0.57 \\
$T_{eff}$ vs. $[M/H]$ &0.11 & 0.09 & 0.02 & 0.11 & 7.11 & 0.14 \\
$M_V$ vs. $[M/H]$ &0.09 & 0.08 & 0.03 & 0.09 & 7.11 & -0.09 \\
$(B-V)$ vs. $[M/H]$ &0.08 & 0.08 & 0.01 & 0.08 & 7.11 & 0.01 \\
$g$ vs. $[M/H]$ &0.07 & 0.06 & 0.02 & 0.07 & 7.11 & 0.10 \\
\hline
\multicolumn{7}{l}{\bf{47 Tuc}} \\
$M_V$ vs. $(B-V)$ & 0.75 & 0.54 & 0.22 & 0.75 & 6.43 & -0.45 \\
$M_V$ vs. $g$ & 0.62 & 0.56 & 0.03 & 0.62 & 6.43 & 0.23 \\
$g$  vs. $T_{eff}$ & 0.56 & -0.01 & 0.03 & 0.56 & 5.75 & 0.76 \\
$(B-V)$  vs. $T_{eff}$ & 0.56 & 0.08 & 0.04 & 0.56 & 6.29 & -0.69 \\
$(B-V)$ vs. $g$ & 0.54 & -0.03 & 0.06 & 0.54 & 6.13 & -0.75 \\
$M_V$ vs. $T_{eff}$ & 0.50 & 0.49 & 0.09 & 0.50 & 6.43 & 0.14 \\
$M_V$ vs. $\xi$ & 0.38 & 0.36 & 0.08 & 0.38 & 6.25 & -0.12 \\
$[\alpha / Fe]$ vs. $\xi$ & 0.34 & 0.32 & 0.11 & 0.34 & 6.43 & -0.13 \\
$T_{eff}$ vs. $\xi$ & 0.29 & 0.28 & 0.13 & 0.28 & 6.43 & 0.07 \\
$[m/H]$ vs. $[\alpha / Fe]$ & 0.23 & -0.09 & 0.01 & 0.23 & 6.36 & -0.57 \\
$g$ vs. $\xi$ & 0.23 & 0.12 & 0.03 & 0.23 & 6.43 & -0.33 \\
$(B-V)$ vs. $\xi$ & 0.22 & 0.14 & 0.06 & 0.22  & 6.43 & 0.28 \\
$M_V$ vs. $[m/H]$ & 0.21 & 0.20 & 0.09 & 0.21 & 6.43 & 0.11 \\
$M_V$ vs. $V_{rot}$& 0.19 & 0.19 & 0.14 & 0.19 & 6.43 & -0.02 \\
$g$ vs. $[m/H]$ & 0.17 & 0.03 & 0.02 & 0.17 & 6.43 & 0.38 \\
\hline
\end{tabular}
\end{table*}

As Table~\ref{ddmic} shows, for the SIMBAD data, MIC identifies statistically significant relationships between $M_V$ and $(B -V)$, $T_{eff}$ and $g$ respectively but not $[M/H]$. Those involving $T_{eff}$ and $(B-V)$ are also more likely to be nonlinear in nature than that with $g$. In fact, there seems to be a general set of relationships between $M_V$, $(B-V)$, $T_{eff}$ and $g$ but not with $[M/H]$. Certainly, this is in line with the Eureqa formulae where the $[M/H]$ dependence is not complex but strictly linear. The MIC results for the 47 Tuc data show the significant relationships found in the SIMBAD data set but also ones involving microturbulence and metallicities as we would expect for PMS stars. Note that there is a weak dependence between $V_{rot}$ and $M_V$ but not between it and any other parameter. The relationships between $M_V$ and $(B_V), g, T_{eff}$ and $\xi$ are also again more likely to be nonlinear in nature.

\subsection{The fundamental plane of elliptical galaxies}
The global properties of elliptical galaxies, such as luminosity, projected velocity dispersion, etc., form a two-dimensional family (\cite{dd87}, hereafter DD87). In particular, an empirical relationship was found by multibivariate statistics between the mean surface brightness, central velocity dispersion, and effective radius of an elliptical galaxy -- the so-called fundamental plane -- which can be employed as a distance indicator, e.g., \cite{dressler87}. This has its physical basis in the virial theorem, 
although there are further structural dependencies exhibited between dwarf and giant ellipticals (\cite{glb93}).

Using Eureqa, we searched the original data set (161 objects) used by DD87 for relationships of the form:

\[
\log(r_e) = f(\log \sigma, <\!\mu\!>, M_g)
\]

\noindent
where $r_e$ is the semimajor axis, $\sigma$ is the velocity dispersion, $<\!\mu\!>$ is the mean surface brightness and $M_g$ is the absolute magnitude in the $r_G$ band. We used a slightly modified set of building blocks from that which we used in the previous section, in that we also allowed for periodic behaviour which could be described in terms of a sine function. This increases the size of search space available, allowing for a wider set of possible relationships, but also, potentially, the computation time. 
We note, however, that have no expectation of periodic behaviour; in fact, we know that it would make no sense in this particular context. Rather part of the experiment is just to see what effect allowing for it in the building blocks might have. We also employed a fitness metric based on the mean absolute error. 

\begin{figure}
\centering
\includegraphics[width=3.5in]{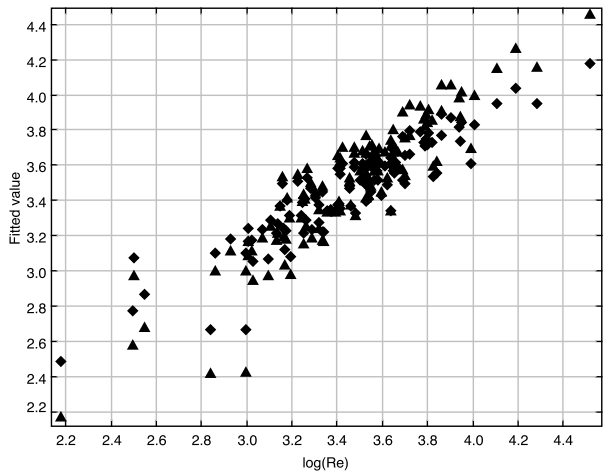} 
\caption{This shows the relative distributions of the original fundamental plane relationship discovered by Djorgovski \& Davis (1987) (diamonds) and that found by Eureqa (triangles). }
\label{fig2}
\end{figure}

The best fit (lowest complexity, highest accuracy) formula was:

\[
\log(r_e) = \log \sigma + 0.271 <\!\mu\!> - 4.09
\]

\noindent
which should be compared with the original relationship reported by DD87 (see also Fig.~\ref{fig2}):

\[
\log(r_e) = 1.39 (\log \sigma + 0.26 <\!\mu\!>) - 6.71
\]
\noindent
The two fits have equivalent accuracies - both give rms errors of 0.157 and correlation coefficients of 0.91. Higher order formulae give even slightly better fits, e.g., $\log(r_e) =  (0.14 \log \sigma - 1)<\!\mu\!> -  M_g$ with 0.142 and 0.92 respectively, but there is a danger that this is overfitting the data, particularly given the small size of the data set. We note also that the sine function was not used.

Given that both relationships are sufficiently similar in form (complexity) and accuracy, the question arises as to which one is correct? This judgement call is beyond the scope of current discovery systems and is where the (human) expert must step in and provide the necessary interpretative knowledge. In this case, the relationships are encoding physical correlations between the size of a galaxy and its effective surface brightness and the luminosity and central velocity dispersion. Even though the form of the expressions is similar, they actually translate into quite different predictions. The DD87 formula gives:

\[
<\!\mu\!> \, \sim \, L^{-\frac{5}{4}} \,\, \mathrm{and} \,\,   D_n \propto \sigma_0^{1.4}<\!\mu\!>^{-0.07}
\]

\noindent
where $D_n$ is the diameter within which the mean surface brightness is $20.75\mu_B$ (\cite{dressler87}), whereas the Eureqa result gives:

\[
<\!\mu\!> \, \sim \, L^{-3} \,\, \mathrm{and} \,\,   D_n \propto \sigma_0<\!\mu\!>^{0.16}
\]

\noindent
implying that more luminous galaxies have much lower surface brightnesses and that the distances to galaxies is less than that actually seen. 

The Eureqa fit makes no use of $M_g$ and the value of MIC for this variable relative 
to  $\log(r_e)$ is the lowest, consistent with a lack of dependency. There is also no indication of any type of bivariate relationship beyond a linear one, although Eureqa finds non-linear multivariate relationships to which the MIC is most likely not too sensitive.

\subsection{Binary classification of light curves}
Determining whether an object belongs to a specified class or not, e.g., a transient detection is a supernova or not, or, alternatively, whether it falls into one of two different (mutually exclusive) classes, such as star or galaxy, is an increasingly common activity in astronomy (note that multi-class classification problems can always be recast as a series of such binary decisions). This is particularly true of survey astronomy where large data volumes and, most recently, real-time data streams require fast, accurate, and reliable classification systems. A variety of techniques have been employed in response, e.g. \cite{flash12}, including decision trees (e,g., \cite{vascon11}), Bayesian networks (e.g., \cite{dubath11}), and support vector machines (SVM, e.g., \cite{beaumont11}), the latter representing the state of the art.

Although it seems somewhat counterintuitive, automated discovery systems can also be used as binary classifiers. With Eureqa, the ``trick'' is formulate the search relationship as:

\[
\mathrm{class} = g(f(x_1, x_2, x_3, \ldots, x_n))
\]

\noindent
where $g$ is either the Heaviside step function or the logistic function, which gives a better search gradient and can be used to produce ROC\footnote{A ROC curve is a graphical plot which summarizes the performance of a classifier over a range of tradeoffs between true positive and false positive errors rates (see Fig.~\ref{fig3}.)} curves for the resulting classification. Eureqa finds a best-fit function, $f$, to the data that will get mapped to a 0 or a 1, depending on whether it is positively or negatively valued (or lies on either side of a specified threshold value, say 0.5, in the case of the logistic function). In other words, it finds an equation for the discriminating hyperplane which separates the two classes in some high-dimensional feature space. This is comparable to what a SVM\footnote{A Support Vector Machine (SVM) is the state-of-the-art binary classification algorithm.} does but with an explicit computation of the mapping into feature space rather than just relying on inner products within it. An advantage of this approach is that the structure of the analytical fit function can also give insight into how the classification works, which is not normally true of other "black box" classifiers, such as neural networks.

The Catalina Real-Time Transient Survey (CRTS; \cite{crts09, djor12, mahabal11}) is the largest open time domain survey currently operating, covering $\sim33000$ deg$^2$ between $-75\degr < {\mathrm Dec} < 75\degr$ (except for within $\sim10 - 15\degr$ of the Galactic plane) to $\sim20$ mag. Light curves of several hundred million objects are available\footnote{http://crts.caltech.edu} with an average of $\sim250$ observations over a 7-year baseline. A common approach to light curve classification is to characterize the light curves through extracted features, such as moments, flux and shape ratios, variability indices, and periodicity measures. Vectors of such features derived from the light curves of known classes of objects are then used as the training sets for particular classifiers.

We have considered three specific binary light curve classification problems using Eureqa: RR Lyrae vs. W UMa, CV vs. blazar, and Type Ia vs. core-collapse supernova. For each case, we compiled data sets of light curves of the appropriate classes of object and derived $\sim\!30$-60-dimensional feature vectors for each object (see Appendix 1 for the full list of features used). These are a combination of the features used by \cite{rb11} and \cite{deb07} and include statistical moments, flux ratios, Stetson J and K variability indices, a quasar-fitting measure and frequency analysis statistics.

We ran a set of 10 4 CPU-hour Eureqa runs (1 hr on a quad-core machine) for each of three cases with each run omitting 10\% of the data (giving training sets that are 90\% of the data set) and the best-fit solution for that run (defined as the least complex which produces the largest number of true positive and negative class attributions) then applied with the omitted data as the validation set so giving us 10x-cross-validation on the resulting solutions. We report our results (see Table~\ref{table2}) in terms of the sum of all the results from the cross-validation runs. A logistic function was used in all cases to map the fitted function to the class variable. The physical interpretation of any relationships identified in these problems is beyond the scope of the present paper and will be addressed elsewhere.

\subsubsection{RR Lyrae vs. W UMa}

\begin{figure}
%\centering
\includegraphics[width=3.55in]{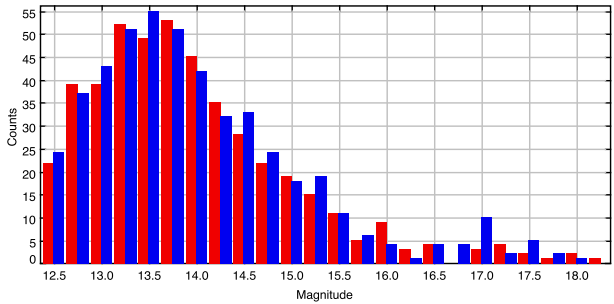} 
\caption{This shows the magnitude distributions of the RR Lyrae (blue) and eclipsing binary (red) data sets used in the binary classification analysis.}
\label{rrwumamag}
\end{figure}

Eclipsing binaries (W UMa) are the predominant contaminant in studies using RR Lyrae as tracers of Galactic structures, e.g., \cite{sesar11}, and therefore being able to distinguish between them would be useful. We extracted CRTS light curves for 482 RR Lyrae and 463 W UMa from SIMBAD and 
the AAVSO International Variable Star Index (\cite{vsx}) %417 RR Lyrae in SDSS Stripe 82 and 222 W UMa from the General Catalog of Variable Stars (GCVS, version 2011-May, \cite{gcvs}) 
obtained from VizieR (\cite{vizier}). The magnitude distribution for both classes of objects are shown in Fig.~\ref{rrwumamag}. Since both classes of object are periodic, we included periodic features in our characterization and used 60-dimensional feature vectors. 

\begin{figure}
\centering
\includegraphics[width=3.5in]{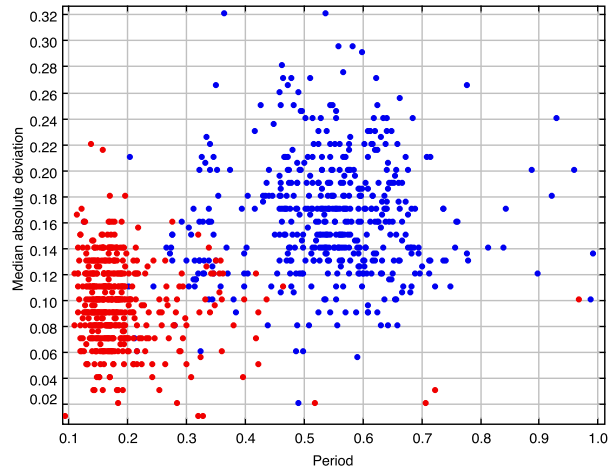}
\caption{This shows the distribution of RR Lyrae (blue) and W UMa (red) stars in the period - median absolute difference plane identified by Eureqa.}
\label{freqmad}
\end{figure}

The overall best-fit formula was:
 \[
f =  278\,x_{24} - \frac{6.63}{x_{10}} - 24
 \]
  
 \noindent 
where $x_{24}$ is the principal period from the Lomb-Scargle periodogram (\cite{lomb76, scargle82}) and $x_{10}$ is the median absolute deviation. The resulting values of class are 1 for RR Lyrae and 0 for W UMa objects. The combined confusion matrix for the best-fit classifying formulae, i.e., summing the individual cross-validation results, is shown in Table~\ref{table2}a and the ROC curve showing the dependencies between the true and false positive classification rates respectively as the logistic function threshold value is varied in Fig.~\ref{fig3}a. It is interesting to note that this is essentially the period-amplitude relation which is used to differentiate between subclasses of RR Lyrae (e.g., \cite{smith11}). Fig.~\ref{freqmad} shows how the two populations are clearly separated in this parameter plane.

MIC measures were calculated for all pairs of features in this feature set. We would expect that significant relationships would be found for pairs of variables having a common basis, e.g., those derived from the Lomb-Scargle periodograms of the light curves or those which measure the fraction of outliers or degree of spread in the light curve, and this was confirmed. The MIC measure also largely correlated with the regression coefficient for these pairs, i.e., those with a high MIC value had a high $r^2$ value as well and vice versa, but one strongly related pair (MIC close to 1) had a very low linear regression ($\sim0.12$). The nonlinearity MIC statistic indicated was also large for this pair and the two features were found to be inversely proportional to each other. This clearly illustrates the power of MIC over traditional bivariate relationship analysis algorithms.

MIC analysis of this feature set -- calculating the MIC measures for all pairs of features -- showed significant relationships between expected pairs of variables, e.g., those derived from the Lomb-Scargle periodograms of the light curves or those which measure the fraction of outliers or degree of spread in the light curve. These largely correlated with the regression coefficient for these pairs but one strongly related pair has a very low linear regression ($\sim0.12$). The nonlinearity MIC statistic indicated such a relationship and the two features were found to be inversely proportional to each other. This clearly illustrates the power of MIC over traditional bivariate relationship analysis algorithms.

Looking for relationships between the class variable for the data set and the features showed a number of significant ($p < 10^{-4}$) pairings. All associations were also deemed to be complex and, with the exception of the median absolute deviation, nonlinear. We'll discuss these in further detail later in the paper.

\subsubsection{CV vs. blazar}

The light curves of cataclysmic variables (CVs) and blazars can be difficult to differentiate as both exhibit aperiodic/quasiperiodic variability with significant (several magnitudes) sudden outbursts. We extracted CRTS light curves for 404 known CVs\footnote{http://nesssi.cacr.caltech.edu/catalina/CVservice/CVtable.html} and 120 Fermi and MOJAVE blazars\footnote{http://nesssi.cacr.caltech.edu/catalina/Blazars/Blazar.html}. Periodic features were omitted in the characterization, giving 25-dimensional feature vectors. The overall best-fit formula was:

\[
f = 140067x_{17}\sin\left(\frac{-0.979}{x_1-1.481}\right) - 264152
\]
\noindent
where $x_1$ is the amplitude and $x_{17}$ is the significance of the chi-squared quasar statistic (\cite{bb11}). The combined confusion matrix for the best-fit classifying formulae is shown in Table~\ref{table2}b and the ROC curve in Fig.~\ref{fig3}b. From the matrix, the classifier is clearly more successful at identifying CVs than blazars. This may reflect stronger class localization for CVs in the feature space than for blazars, i.e., the distribution of CVs in the feature space is more compact and therefore a discriminating (bounding) hyperplane is more easily defined than for blazers. However, it is more likely due  to the 10:3 population ratio of CVs and blazers in the data set and a learning bias -- the so-called test distribution effect (\cite{weiss}) -- that this has created in the classifier, i.e., with more exposure to CVs, the algorithm has preferentially evolved to classify them. We defer further discussion of this issue to section 5.

MIC analysis of the feature set again shows a number of expected significant relationships, i.e., flux ratios and quasar statistics, although the correlation with the respective regression coefficients is much less than in the RR Lyrae vs W UMa case, which may be related to the lack of periodic features. The relationships also tend to be nonlinear but monotonic. In terms of associations with the class variable, only three significant features were found with no clear indication of nonlinearity or non-monotonicity.

\subsubsection{SN Ia vs. core-collapse SN}

\begin{figure*}
\centering
\includegraphics[width=7in]{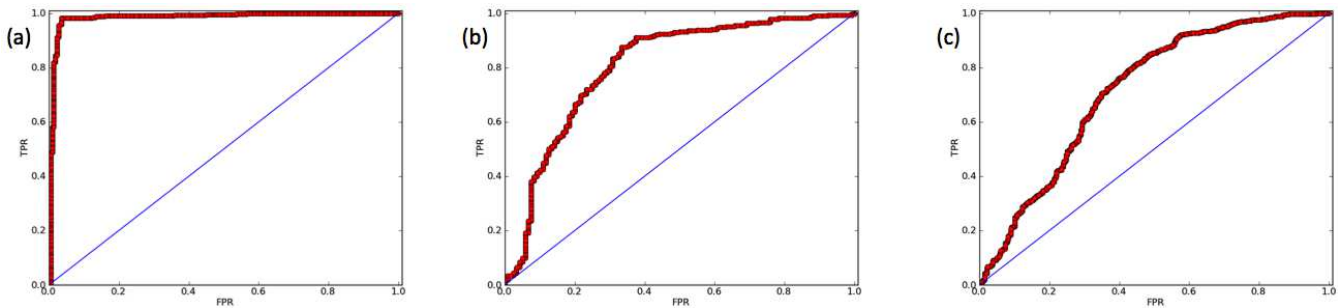} 
\caption{This shows the ROC curves -- true positive rate (TPR) vs. false positive rate (FPR) as the discrimination threshold is varied -- for the three binary classification cases using Eureqa: (a) RR Lyrae vs. W UMa; (b) CV vs. blazar; and (c) SN Ia vs. core-collapse SNe. The best possible classifier would yield a point in the upper left corner (0,1) of the plot. The area under the curve (AUC) is an accepted performance metric for a ROC curve and related to the Mann-Whitney $U$ statistic. The line $y = x$ represents the scenario of randomly guessing the class.}
\label{fig3}
\end{figure*}

Spectroscopic confirmation of supernovae candidates can be resource intensive and becomes intractable with the increasingly large numbers expected from the next generation of wide-field surveys, e.g., a few hundred thousand from Pan-STARRS and LSST. The Supernova Photometric Classification Challenge (SPCC; \cite{snc10}) aimed to improve the state-of-the-art of supernova classification algorithms based solely on photometric data, and, in particular, separating out SNe Type Ia, which is important for cosmological studies. We tested our methodology on a set of 836 SNe Ia and 427 core-collapse SNe (Ib, Ic, IIn, IIp) light curves from the SPCC data set. Again, since we do not believe these to be periodic, we used only non-periodic features to characterize the light curves, giving 25-dimensional feature vectors. The overall best-fit formula was:

\[
f = \frac{x_{18} - 22.9}{x_{15} + 0.21x_{18}} - x_{13} - x_{10}
\]
\noindent
where $x_{10}$ is the median absolute deviation, $x_{13}$ is the percentage difference between the extremum flux and the median, $x_{15}$ is the chi-squared quasar statistic, and $x_{18}$ is the significance of the chi-squared non-quasar statistic respectively. The combined confusion matrix for the best-fit classifying formulae is shown in Table~\ref{table2}c and the corresponding ROC curve in Fig.~\ref{fig3}c. The matrix again shows a strong classification bias for the more numerous class, although this time the population ratio is only $\sim$2:1.

The MIC results show expected relationships between flux ratios and measures of variability, all of which are mainly linear, monotonic, and in line with the respective regression coefficient results. More interestingly, though, is that there are no really significant associations between the class variable and the features -- the most significant, the ratio between the (95th - 5th) flux percentile and the median, is only significant at the $\sim$3\% level. This may indicate that the conventional set of features used to characterize light curves are inappropriate for those of supernovae, which would also explain the poor performance of the classifier - a clear discriminating hyperplane cannot be defined in this feature space.

\begin{table}
\caption{The combined best-fit confusion matrices for the three binary classification cases using Eureqa and 10x-cross-validation. The results are the sums of each cross-validation run.}
\label{table2}
\centering
\begin{tabular}{lll}
\hline
(a)  & RR Lyrae & W UMa \\
\hline
RR Lyrae & 464 (96.3\%) & 18 (3.7\%) \\
W UMa & 7 (2.5\%) & 456 (98.5\%) \\
\hline
\hline
(b) & CV & Blazar \\
\hline
CV & 368 (91.1\%) & 36 (8.9\%) \\
Blazar & 45 (37.5\%) & 75 (62.5\%) \\
\hline
\hline
(c) & SNe Ia & CC SNe \\
\hline
SNe Ia & 773 (92.5\%) & 63 (7.5\%) \\
CC SNe & 250 (58.6\%) & 177 (41.4\%) \\
\hline
\end{tabular}
\end{table}

\section{Feature selection}

\subsection{Posterior feature selection}
In examining data mining systems, it is often worth asking whether a successful outcome is due to the power of the particular algorithm under consideration or due to a comprehensive training data set being used, with which any algorithm worth its salt would achieve good results. (Alternatively, a poor result with an otherwise excellent algorithm may be due to a limited training set). One way to answer this is to consider which features in the data set are employed by the algorithm and ask whether the features selected show any degree of sense -- do they provide additional insight into the data set -- or should we regard them purely as phenomenological selections that just happen to give good results? This is particularly so when only a subset of all the available features actually end up being used, i.e., there is some degree of feature selection present in the process, whether explicit or implicit (embedded), as happens with evolutionary-based algorithms and the decision tree work we have compared Eureqa against.  

The MIC statistics already give some handle on the relative importance of different bivariate relationships within the feature space and of particular features relative to the class variable in the classification examples. However, we would also like to be able to consider larger multivariate subsets of features, both for feature ranking according to some metric and to identify the optimal subset of features that characterizes the problem. We have considered two further specific feature selection techniques to compare against the results of Eureqa, MIC and the decision trees
and determine whether there is any consistency in the features used by the different techniques: consensus would imply that the shared features are relevant to understanding the problem under consideration.

\subsubsection{Sequential backward ranking}
Sequential backward ranking (SBR) is an unsupervised feature selection method based on the entropy measure that aims to progressively reduce the dimension of a data set in an optimal fashion, i.e., at each stage, the reduced data set represents the best approximation to the original. It works thus:

\begin{enumerate}
\item Start with a full feature set $\cal F$ which characterizes a data set
\item For each feature, $f \in \cal F$, define a set of subsets, $\{{\cal F}_f\}$, such that: ${\cal F}_f = {\cal F} - f$ 
\item Select the feature $f_m$ which maximizes the quantity: ${\cal S}({\cal F}) - {\cal S}({\cal F}_{f_m})$ where ${\cal S}({\cal F})$ is the Shannon entropy (see below) of the feature set $\cal F$
\item Update $\cal F$ such that: ${\cal F} = {\cal F} - f_m$
\item Repeat steps (ii) - (iv) until there is only one feature left
\end{enumerate}

\noindent
The output is an ordered list of features in descending order of their entropy contribution or their significance. A supervised version can also be constructed by replacing the constraint in step (iii) with minimizing the classification error between that for ${\cal F}$ and ${\cal F}_{f_m}$.

In order to apply this technique, we must first define and evaluate the Shannon entropy of a feature set.
Traditional estimators of the Shannon entropy, $H(X)$ of a multivariate data set, $X = \{X_1, X_2, \ldots, X_n\}$, require knowledge of the joint probability distribution of all the $X_n$:

\[
H(X_1, \ldots, X_n) = - \sum_{x_1} \ldots \sum_{x_n} P(x_1, \ldots, x_n) \log \left[ P(x_1, \ldots, x_n) \right]
\]

\noindent
which is usually a fairly intractable problem. However, \cite{kl87} provide an alternative estimator based on the distance to the k$^{th}$-nearest neighbour:

\[
H(X_1, \ldots, X_n) = -\psi(k) + \psi(n) + \log c_d + \frac{d}{n} \sum_{i=1}^{n} \log(\epsilon_i)
\]

\noindent
where $\psi$ is the digamma function ($\psi(x) = \Gamma'(x) / \Gamma(x)$), $c_d$ is the volume of the $d$-dimensional unit ball ($c_d = \pi^{d / 2} / \Gamma(1 + d/2)$) and $\epsilon_i$ is twice the distance from $x_i$ to its k$^{th}$-nearest neighbour respectively. The error on the estimate is typically $\sim k/N$ or $\sim k/N\log(N/k)$.

\subsubsection{Minimum-redundancy-maximum-relevance}
In feature selection, it has been recognized that the combinations of individually good features do not necessarily lead to good overall performance, i.e., the $m$ best features are not the best $m$ features (e.g., \cite{cover74}). 
One way to tackle this is to consider simultaneously the relevance -- the average mutual information between a set of features and a classification variable -- and the redundancy -- the average mutual information between pairs of features -- of a feature set. \cite{mrmr} proposed such a criterion (minimum-redundancy-maximum-relevance (mRMR)):

\[
\max_S \left[ \frac{1}{|S|} \sum_{f_i \in S} MI(f_i;c) - \frac{1}{|S|^2} \sum_{f_i,f_j \in S} MI(f_i; f_j) \right]
\]

\noindent
where the feature set $S$ has individual features $f_i$, $c$ is the classification variable and $MI$ the mutual information (see eqn.~(1)) respectively. This approximates maximizing the mutual information between the joint distribution of the selected features and the classification variable but in terms of bivariate quantities and not much harder to deal with multivariate ones.  Note that mRMR employs data discretization as a preprocessing step for continuous data since it is often difficult to compute the integral form of eqn.~(1) in a continuous space with limited numbers of samples.

\subsubsection{Feature comparison}

Table~\ref{features} gives the (ordered) lists of features selected by each method -- MIC, Eureqa, decision tree, SBR, and mRMR -- for the three data sets used in the binary classification problems. We did not consider the Hertzsprung-Russell diagram or fundamental plane of elliptical galaxies data sets since these both involved too few variables to show any differences between the methods. The MIC results are those deemed to be statistically significant at the 10$^{-3}$ level relative to the classification variable (see section 2.1 for details). Similarly, the mRMR results are also relative to the classification variable. Feature data for the mRMR algorithm was discretized into three states at the positions $\mu \pm \sigma$ (where $\mu$ is the mean and $\sigma$ the standard deviation respectively) such that it takes -1 if it is less than $\mu - \sigma$, 1 if larger than $\mu + \sigma$ , and 0 if otherwise. The SBR and mRMR results also just list the top five features in each case. Finally, the Eureqa and decision tree entries list the variables used without any implied ranking.

The disparate nature of the rankings - ordered and unordered, different numbers of variables - makes any formal quantitative analysis, such as ranking aggregation, difficult. However, there a number of general comparisons that can be made. The features employed by Eureqa and decision trees are generally different -- they only share one feature in each of the three problems -- and decision trees are less parsimonious with more features. A similar lack of commonality is shown between MIC and both SBR and mRMR, although SBR and mRMR show a marginally stronger degree of overlap, which should not be that surprising since they are both rely directly on entropy-related measures. Table~\ref{relative} gives the relative fractions of the features selected by Eureqa and decision trees that are also identified by MIC, SBR and mRMR respectively. This suggests that there is more agreement between Eureqa and the three explicit feature selection methods than between decision trees and the same techniques, although none of them display any particularly strong association.
 
The differing nature of the classes of object in the three experiments leads one to expect that specific features or types of features would be selected in each and this does seem to be the case. The RR Lyrae/W UMa results include periodic measures for all methods except mRMR, reflecting the period-amplitude relationship, and the CV/Blazar results include either the QSO or non-QSO statistic for all methods. More interestingly, the QSO statistics are also selected by Eureqa, SBR and mRMR as discriminating features in the SNe data set (the other features selected are not common across the methods). These statistics measure the applicability of a damped random walk model to a light curve versus it exhibiting temporally uncorrelated variability. Although neither behavior is shown in either type of supernova light curve, both exhibiting a brightening and then decaying pattern with additional features in core-collapse supernovae, there must be some further information inherent in the light curves to which these statistics are sensitive.
 
\cite{graczyk} propose that eclipsing binaries can be identified in large photometric surveys based on the skew and kurtosis of their light curves. MIC finds some degree of relationship between the skew and kurtosis but not a nonlinear one and a stronger nonlinear dependence between the class type and the skew than the kurtosis (the dependence of which is actually not statistically significant). mRMR identifies both skew and kurtosis, however, as significant features. Neither are flagged by Eureqa or decision trees, although in the former case, the "survivability" of the period - MAD solution dominates that of other possible relationships. Restricting Eureqa to non-periodic features gives a set of formulae all dependent on the skew and variously the kurtosis and percentile ratios. A viable Eureqa-based feature selection strategy, particularly for feature-rich data sets, might therefore be to progressively restrict the set of features that are considered in any single iteration. 
 
It is also worth noting which features are not selected at all or only once by one method: flux ratios and the Stetson K variability index. These statistics can be broadly thought of as quantitative measures of the shape of the light curve and there is indeed little discrimination to be found in the shapes of the three binary categories of object alone, e.g., although RR Lyrae AB and W UMa are relatively easily distinguished from their phased light curves, RR Lyrae C are not. CV and blazar light curves are similarly not easily separable, particularly when sparsely and irregularly sampled such as the CRTS light curves. SN Ia and core-collapse SNe can be differentiated if enough of the light curve has been sampled but this is not necessarily the case with many of the examples in the SPCC.

This suggests that the current features which aim to capture the shape of a light curve are neither robust enough in the presence of noisy inhomogeneous data nor do they capture enough information to act as significant discriminators. Clearly further research in this area would be extremely beneficial to the next generation of time domain surveys.

\begin{table*}
\caption{The features selected by the various feature selection methods discussed in the text. For MIC, SBR and mRMR, the lists are in descending order of feature significance (indicated by $^{\ast}$) . The features to which the variables correspond are described in Table~\ref{table3}.}
\label{features}
\centering
\begin{tabular}{llll}
\hline
Method & RR Lyrae/W UMa & CV / Blazar & SNIa/CCSNe \\
\hline
MIC$^{\ast}$ & $x_{24}, x_{26}, x_{34}, x_{14}, x_{28}, x_{37}, x_{19}, x_{39}$ & $x_{17}, x_{22}, x_{21} $&  -- \\
Eureqa & $x_{10},x_{24} $ & $x_1, x_{17}$ & $x_{13}, x_{15}, x_{18}$ \\
Decision tree & $x_{10}, x_{14}, x_{22}, x_{21}, x_{35}, x_{62}$ & $x_1, x_2, x_6, x_9, x_{15}, x_{19}, x_{22} $ & $x_5, x_{10}, x_{12}, x_{13}, x_{14}$\\
SBR$^{\ast}$  & $x_9, x_{52}, x_{43}, x_{20}, x_{32}$ & $x_{18}, x_{16}, x_{15}, x_{22}, x_{20}$ & $x_{22}, x_{18}, x_{16}, x_{15}, x_9 $ \\
mRMR$^{\ast}$  & $x_{9}, x_{20}, x_{11}, x_{19}, x_{13}$ & $x_2, x_{17}, x_{15}, x_{12}, x_7$ & $x_1, x_2, x_{22}, x_{15}, x_8$ \\
\hline\\
\end{tabular}
\end{table*}

\begin{table*}
\caption{The relative fractions of shared features between those identified by Eureqa or decision trees and those that the three ordered algorithms have provided.}
\label{relative}
\begin{tabular}{lccc}
\hline
Method & RR Lyrae/W UMa & CV / Blazar & SNIa/CCSNe \\
(MIC/SBR/mRMR)	& (8/5/5) & (3/5/5) & (0/5/5) \\
\hline
Eureqa & 50\%/0\%/0\% & 50\%/0\%/50\% & 0\%/67\%/33\% \\
Decision tree & 16\%/0\%/0\% & 14\%/28\%/14\% & 0\%/0\%/0\% \\
\hline
\end{tabular}
\end{table*}

\section{Discussion}
The results in the previous sections show that automated discovery systems of relationships can identify and characterize physically meaningful structure in data. The fact that known relationships in the Hertzsprung-Russell diagram, the fundamental plane of elliptical galaxies and the period-amplitude plane of RR Lyrae stars can be automatically recovered is very encouraging, particularly as more complex and accurate (with smaller errors) expressions are possible. The fitness metrics, however, provide a good balance between accuracy and parsimony, ensuring high quality general hypotheses. It should also be noted that the discovery process consists not only of identifying the best functional expressions but also the most relevant subset of variables. There are, of course, other specific feature selection algorithms, such as those mentioned in section 4, that could be have been applied to the data sets prior to the application of our methods as a preprocessing step.

Perhaps one of the more surprising applications of these systems is as part of efficient binary classifiers, particularly as it has been said that we should not expect a lightly parameterized form for mapping between feature space and class space (Richards 2011, private communication). To get some idea of how competitive this approach is, we can compare directly with the results of \cite{dt12} who have applied C4.5 decision trees using the Gini diversity index as the splitting criterion to the same data sets as in section 3. Table~\ref{dtcompare} gives the relative performances of the two approaches in terms of purity - the fraction of true classifications recovered out of all objects assigned to that class - and efficiency - the fraction of true classifications recovered out of all objects actually belonging to that class. For example from Table~\ref{table2}, 464 true RR Lyrae are recovered, 471 (464 + 7) objects are assigned a class of RR Lyrae, and there are 482 (464 + 18)  RR Lyrae in the data set - this gives a purity of 464 /  471 (98\%) and an efficiency of 464 / 482 (96\%). It can be seen that for four of the classes, the Eureqa-based approach performs as well as the decision tree one, particularly in terms of efficiency (completeness). For additional comparison, the best results reported in SPCC were 96\% efficiency and 79\% purity for classifying SNe Ia (\cite{snc10}). However, as expected, it does not perform so well with the two minority class populations.

{\em Imbalanced} data sets, such as the CV/blazar and SNIa/CCSN examples, may reflect natural class distributions -- one type of object is just more common than the other -- or may be the result of parameter/feature space sampling -- observations are probing regions preferentially occupied by one class, even if the overall population sizes are similar. We chose to use as much data as possible in both the CV/blazar and SNIa/CCSN cases, which probably involves a mixture of both these effects. With such data sets, minority class examples are classified incorrectly much more often than majority class examples (\cite{weiss}), as we found in section 3. Determining what the correct distribution is for a learning algorithm in this context is an active area of research in machine learning (see \cite{chawla} for an overview). Some practitioners believe that the naturally occurring marginal class distribution should be used so that new examples will be classified using a model built from the same underlying distribution. Others feel that the training set should contain an increased percentage of minority class examples or the induced classifier will not classify minority class examples well. \cite{weiss} show that the choice of training distribution can depend on the performance measure used with the natural distribution for predictive accuracy (confusion matrices) and a balanced distribution for ROC curves respectively. Although, \cite{chawla} argues that predictive accuracy may be an inappropriate performance measure for imbalanced data sets.
Our results are certainly consistent with all these findings. 

\begin{table}
\caption{The overall success rates for the Eureqa-based classifiers and the decision trees of Donalek et al. (2012)}
\label{dtcompare}
\centering
\begin{tabular}{lcccc}
\hline
Data set & \multicolumn{2}{c}{Eureqa} & \multicolumn{2}{c}{Decision tree} \\
& Purity & Efficiency & Purity & Efficiency \\
\hline
RR Lyrae & 98\% & 96\%&  95\% & 95\% \\
W UMa & 97\% & 99\% & 96\% & 96\% \\
CV & 89\% & 91\% & 92\% & 92\% \\
Blazar & 68\% & 63\% & 87\% & 83\%\\
SN Ia & 76\% & 93\% & 90\%& 96\%\\
CC SN & 74\% & 41\% & 92\% & 80\%\\
\hline
\end{tabular}
\end{table}

Finally, it is worth considering the limitations of automated discovery systems. Eureqa assumes that relationships must be expressible as invariant (conserved) quantities in a partial differential metric space. However, this would not necessarily be true for systems that might be exhibiting fractal behaviour, such as scale dependency in the correlation properties of the large-scale distribution of galaxies (e.g., \cite{joyce05}), or chaotic or stochastic activity, such as in accretion discs (e.g., \cite{karak10}). It is computationally expensive and more than linearly so as the size of the search space is increased with the number of building blocks employed in search formulae.  It can also suffer from the general limitations of evolutionary algorithms, requiring time to move out of local minima and the nature of the fitness landscape being unclear so it is difficult to determine how well the algorithm might be doing. Although there is no guarantee that it will find a good solution nor that this will be the optimum, the results we have shown demonstrate that it is a useful technique to consider.

MIC and its associated statistics have fewer assumptions in just looking for and broadly characterizing bivariate relationships through their maximal mutual information -- note they are not directly related to mutual information as they perform well in situations where other direct mutual information-based measures do not (\cite{mic11}). Ideally the MIC algorithm would optimize over all possible grid partitions of a data set but the computational expense is avoided with a dynamic programming approach that appears to approximate well in most cases. It is unclear, however, how well these perform in the presence of outliers or how large data sets need to be for stable estimates. Probably the biggest current limitation to MIC is its bivariate nature --  generalizations to higher dimensions are necessary to search for multivariate relationships (these will not necessarily show in a two-dimensional projection) but this comes at the additional expense of both finding an optimal hypergrid partitioning of the data set and also using multivariate mutual information which is a poorly understood concept.

\section{Conclusions}
In this paper we have demonstrated that automated discovery systems can uncover significant (non-trivial) relationships in high-dimensional complex data parameter spaces. As with any relationship found in data, whether by an automated system or a human, these may or may not have a physical meaning or cause -- correlation does not imply causation -- and they may be due to some incidental properties of a given data set. The interpretation and evaluation of their possible physical significance remains in the hands of a human scientist.

Whilst the ones we have shown may not be the most scientifically exciting, being more for illustrative purposes than anything, we should bear in mind that astronomy is an already relationship-rich science. Many of these are expected or predictable associations, given what we already understand about the nature of (astro)physical systems. In contrast, systems biology and similar sciences, wherein lie the origins of these automated discovery techniques, are relationship-poor and there is potentially more upfront impact to be had by applying them in that particular context. Astronomy perhaps stands to benefit more from them as discovery filters, tackling the curse of dimensionality of high-dimensional parameter spaces and reducing the number of relationships to be examined to only the most significant, than as {\em ab initio} discovery engines.

Although these systems represent the cutting-edge of currently applicable tools, this is very much an initial entry point for their application to astronomy. Such tools will very likely become both more powerful and also more prevalent with time, given the data challenges all sciences are facing. Expanded abilities such as not just relying on brute-force searches of feature spaces but being able to incorporate domain knowledge, both as additional features, e.g, distance to nearest galaxy for supernovae, or as rulesets,
and make inferences leading to more interesting discoveries are active areas of research.  

The importance of such approaches as we are faced with ever more parameter/feature rich data sets cannot be underestimated. In particular, the possibilities of high-dimensional scientific relationships, particularly those that do not necessarily reveal themselves in lower dimension representations, can only really be investigated using automated discovery techniques which are (relatively) unconstrained in their exploration of parameter space. These tools promise to cherry pick the higher hanging fruit of LSST, SKA and future surveys.

\section*{Acknowledgments}

We thank Hod Lipson and Michael Schmidt for useful discussions and their kind assistance with the Eureqa software. We also thank the anonymous referee for their useful comments which helped improve this paper.
 
This work was supported in part by the NSF grants AST-0909182 and IIS-1118041, by the W. M. Keck Institute for Space Studies, and by the U.S. Virtual Astronomical Observatory, itself supported by the NSF grant AST-0834235. 

This research has made use of data obtained from or software provided by the US Virtual Astronomical Observatory, which is sponsored by the National Science Foundation and the National Aeronautics and Space Administration.

This research has made use of the SIMBAD database, operated at CDS, Strasbourg, France, and the International Variable Star Index (VSX) database, operated at AAVSO, Cambridge, Massachusetts, USA.

Funding for SDSS-III has been provided by the Alfred P. Sloan Foundation, the Participating Institutions, the National Science Foundation, and the U.S. Department of Energy Office of Science. The SDSS-III web site is http://www.sdss3.org/.

SDSS-III is managed by the Astrophysical Research Consortium for the Participating Institutions of the SDSS-III Collaboration including the University of Arizona, the Brazilian Participation Group, Brookhaven National Laboratory, University of Cambridge, Carnegie Mellon University, University of Florida, the French Participation Group, the German Participation Group, Harvard University, the Instituto de Astrofisica de Canarias, the Michigan State/Notre Dame/JINA Participation Group, Johns Hopkins University, Lawrence Berkeley National Laboratory, Max Planck Institute for Astrophysics, Max Planck Institute for Extraterrestrial Physics, New Mexico State University, New York University, Ohio State University, Pennsylvania State University, University of Portsmouth, Princeton University, the Spanish Participation Group, University of Tokyo, University of Utah, Vanderbilt University, University of Virginia, University of Washington, and Yale University.

\appendix

\section[]{Characterizing features}

A variety of statistically- and morphologically-based features have been used to characterize light curves in the literature, e.g., \cite{rb11}, \cite{deb07}. Table~\ref{table3} summarizes the statistics that we have used in this analysis.

\begin{table*}
\caption{This table describes the features used to characterize the light curves used in this analysis.}
\label{table3}
\begin{tabular}{lll}
\hline
Name & Variables & Description \\
\hline
Amplitude & $x_1$ & Half the difference between the minimum and maximum magnitudes \\
Beyond 1 std & $x_2$ &  The percentage of points beyond one standard deviation from the weighted mean
 \\
Flux percentile ratio (60 - 40) & $x_3$ &  The ratio of flux percentiles: (60th - 40th) to (95th - 5th)\\
Flux percentile ratio (67.5 - 32.5) & $x_4$ & The ratio of flux percentiles: (67.5th - 32.5th) to (95th - 5th)\\
Flux percentile ratio (75 - 25) & $x_5$ &  The ratio of flux percentiles: (75th - 25th) to (95th - 5th)\\
Flux percentile ratio (82.5 - 17.5) & $x_6$ &  The ratio of flux percentiles: (82.5th - 17.5th) to (95th - 5th)\\
Flux percentile ratio (90 - 10) & $x_7$ & The ratio of flux percentiles: (90th - 10th) to (95th - 5th)\\
Linear trend & $x_8$ &  The slope of a linear fit to the light curve \\
Maximum slope &$x_9$ &  The maximum absolute flux slope between two consecutive observations \\
Median absolute deviation & $x_{10}$ & The median discrepancy of the fluxes from the median flux\\
Median buffer range percentage & $x_{11}$ &  The percentage of fluxes within 10\% of the amplitude from the median \\
Pair slope trend & $x_{12}$ &  The percentage of the last 30 pairs of consecutive flux measurements that have a positive slope \\
Percent amplitude & $x_{13}$ &  The largest percentage difference between either the maximum or minimum flux and the median\\
Percent difference flux percentile & $x_{14}$ &  The ratio of the (95th - 5th) flux percentile to the median flux \\
QSO & $x_{15} - x_{18}$ & The chisq/qso and chisq/non-qso statistics and their significance levels from the quasar \\ & & (non-)variability metric of \cite{bb11}\\
Skew & $x_{19}$ & The skew of the magnitudes\\
Small kurtosis & $x_{20}$ &  The kurtosis of the magitudes\\
Standard deviation & $x_{21}$ & The standard deviation of the magnitudes\\
Stetson J & $x_{22}$ & The Welch-Stetson J variability index with an exponential weighting scheme\\
Stetson K & $x_{23}$ &  The Welch-Stetson K variability index \\
Lomb-Scargle peaks & $x_{24} - x_{33}$ &  The periods and false-peak detection probabilities of the top 5 peaks in the Lomb-Scargle\\
& &  periodogram of the light curve\\
Frequency parameters & $x_{34} - x_{62}$ &  The frequency analysis statistics described in \cite{deb07}: the slope of the linear\\
& & trend, the 3 prime frequencies and their first four harmonics (amplitude and phase for each) \\
 & & and the ratio of the variances of the light curve after and before subtraction of a harmonic fit \\
 & & with the first frequency.\\
\hline
\end{tabular}
\end{table*}

\bsp

\label{lastpage}

\end{document}